# A micromechanics-based analytical solution for the effective thermal conductivity of composites with orthotropic matrices and interfacial thermal resistance

Sangryun Lee[1,†], Jinyeop Lee[2,†], Byungki Ryu[3] and Seunghwa Ryu[1,*]

**Affiliations**

[1]Department of Mechanical Engineering and [2]Department of Mathematical Sciences, Korea Advanced Institute of Science and Technology (KAIST), 291 Daehak-ro, Yuseong-gu, Daejeon 34141, Republic of Korea

[3]Thermoelectric Conversion Research Center, Korea Electrotechnology Research Institute (KERI), Changwon-si, Gyoengsangnam-do, 51543, Republic of Korea

[†]These authors contributed equally to this work.

[*]Corresponding author e-mail: ryush@kaist.ac.kr

**Abstract**

We obtained an analytical solution for the effective thermal conductivity of composites composed of orthotropic matrices and spherical inhomogeneities with interfacial thermal resistance using a micromechanics-based homogenization. We derived the closed form of a modified Eshelby tensor as a function of the interfacial thermal resistance. We then predicted the heat flux of a single inhomogeneity in the infinite media based on the modified Eshelby tensor, which was validated against the numerical results obtained from the finite element analysis. Based on the modified Eshelby tensor and the localization tensor accounting for the interfacial resistance, we derived an analytical expression for the effective thermal conductivity tensor for the composites by a mean-field approach called the Mori-Tanaka method. Our analytical prediction matched very well with the effective thermal conductivity obtained from finite element analysis with up to 10% inhomogeneity volume fraction.

## Introduction

Due to their exceptional effective thermal conductivity, reinforced composites are widely used in various applications, such as electronic devices[1,2], thermal energy storage[3,4], and thermoelectric devices[5]. For example, the thermal conductivities of epoxy reinforced with silicon[6], graphite[7,8], or aluminium nitride[9] are significantly improved compared to pure epoxy. These highly conductive composites have been investigated for use as efficient thermal heat sinks in the electronic packaging design[10-12]. Meanwhile, composites made of low conductive materials can serve as good insulators[13,14]. To efficiently design and use these composites, it is essential to understand and be able to predict the effective thermal conductivity of the composites.

Numerous previous studies have focused on the effective material properties of composites based on numerical simulations and theories[15-26]. The effective thermal conductivities of macroscale composites have been computed by finite element analysis (FEA) as a function of the shape and orientation of their inhomogeneities[15,16,27]. Molecular dynamics simulations were used to predict the effective thermal conductivities of composites possessing nanoscale inhomogeneities[23,26]. However, multiple calculations are required to obtain statistically meaningful values of effective conductivity by changing the positions, numbers, and the orientation distributions of the inhomogeneities in a simulation cell[16,28]. For example, when the thermal conductivity of a spherical inhomogeneity is twenty time larger than that of a matrix, an FEA study based on a representative volume element (RVE) containing ten inhomogeneities showed that the standard deviation of thermal conductivity became as large as 10% of the predicted thermal conductivity[16]. Sufficiently large RVE calculations result in smaller statistical errors, but they consume significant computational resources.

Alternatively, homogenization theories have been applied to predict the effective thermal conductivity of reinforced composites with a relatively low (< 20%) volume fraction of inhomogeneities[4,21,22,24,25]. In a micromechanics-based approaches to elasticity, the effective stiffness of a composite can be computed when the Eshelby tensor of a given inclusion is known[29]. The Eshelby tensor relates eigenstrain and constrained strain in the single inclusion problem, and can be applied to solve inhomogeneity problems by considering the equivalent eigenstrain. The Eshelby tensor has been theoretically calculated for a variety of inclusion shapes, including prolate spheroids, oblate spheroids, spheres, and cylinders in the case of an isotropic matrix[30]. The eigen-intensity problem in heat conduction is mathematically analogous to the eigenstrain problem in elasticity[22,30] (See **Supplementary Table 1**). Hence, the Eshelby tensor is constant for the ellipsoidal inclusion in heat conduction in the absence of an interfacial resistance, as it is constant for analogous elasticity problems. The effective thermal conductivity of the composites with multiple inhomogeneities can be expressed in terms of the Eshelby tensor and the localization tensor (analogous to the stress concentration tensor in elasticity), based on homogenization approaches such as Mori-Tanaka or self-consistent methods[21,22,24,25].

Theoretical studies for the isotropic matrix and ellipsoidal inclusion without interfacial thermal resistance have well been established[19,21,30]. However, in realistic composites, interfacial thermal resistance arises from many reasons, such as relative roughness, lattice mismatch, and poor chemical or mechanical adhesion[31,32], which results in a temperature jump across the interface. Such interfacial thermal resistance, the so-called Kapitza resistance, reduces the thermal conductivity of the composites[17,20,33]. Another complexity arises because the thermal conductivity of a matrix is anisotropic, i.e., thermal conductivity changes with the direction of the heat flux applied to matrix. For example, single-crystal metals and ceramics in a tetragonal or hexagonal crystal lattice have transversally isotropic thermal conductivity where the conductivities along two directions are identical but different to the conductivity along the other direction[34-36]. Polymer matrices synthesized through extrusion and drawing processes can also have transversally isotropic conductivity because the polymer chains are aligned along one direction[37]. They have different axial and lateral physical properties depending on processing conditions, such as the processing temperature and extrusion rate[37-39]. Orthorhombic crystals such as cementite[40], titanium alloy[41], and tin selenide (SnSe)[42] have three independent material constants in the thermal conductivity tensor[43]. Crystal families with less symmetry can have non-zero off-diagonal components in the thermal conductivity tensor, i.e., the direction of heat flow may not be exactly same as the direction of temperature gradient. There are several experimental studies on the thermal conductivity of composite materials with anisotropic matrices[44,45]. However, existing theoretical studies consider either anisotropic matrices with zero interfacial resistance[46], or isotropic matrices with finite interfacial resistance[22]. To the best of the authors' knowledge, there exists no theoretical study simultaneously considering the anisotropy of matrix and the interfacial thermal resistance.

In this work, we derive, for the first time, an analytical expression for the effective thermal conductivity of composites composed of spherical inhomogeneities and orthotropic matrices by accounting for the effect of interfacial thermal resistance. In the first half of the remainder of this paper, we obtain the analytical expression of the Eshelby tensor for an eigen-intensity problem when the matrix is an orthotropic material in the absence of the interfacial resistance. Based on the Eshelby tensor, we compute the heat flux within a single inhomogeneity and the effective modulus of composite. In the second half, we obtain a modified Eshelby tensor that accounts for the interfacial thermal resistance. For the single inhomogeneity problem, theoretical predictions of the heat flux within the inhomogeneity under external heat flux, and the amount of the temperature jump at the interface, are validated against numerical calculations based on FEA for a wide range of interfacial thermal resistances. We then apply a micromechanics-based homogenization method to derive an analytical solution of the effective thermal conductivities along three axes for the particle-reinforced composites with orthotropic matrices. The effective thermal conductivity prediction correctly converges to that of a porous matrix at the infinite interfacial resistance limit and that of perfect interface solution in the zero interfacial resistance limit. We show that our analytical prediction matches very well with the FEA results for an RVE of particle-reinforced composites.

## Results and Discussion

### Effective Thermal Conductivity in the Absence of Interfacial Resistance

In the steady state, the governing equation for heat conduction is written as,

$$K_{0ij}\frac{\partial}{\partial x_i}\left(\frac{\partial T(\boldsymbol{x})}{\partial x_j}\right)+g(\boldsymbol{x})=0 \tag{1}$$

where $g$ is a heat source, $T$ is temperature field, and $\boldsymbol{K_0}$ is a symmetric second order thermal conductivity tensor. The repeated small letter indices represent the dummy indices that imply summation over all the values from 1 to 3. The Green's function $G(\boldsymbol{x}-\boldsymbol{y})$ in the steady state heat conduction equation is defined as the temperature field at a position $\boldsymbol{x}$ in the presence of a unit heat source at another position $\boldsymbol{y}$ in an infinite medium,

$$K_{0ij}\frac{\partial^2 G(\boldsymbol{x}-\boldsymbol{y})}{\partial x_i \partial x_j}+\delta(\boldsymbol{x}-\boldsymbol{y})=0.. \tag{2}$$

By solving the equation, the Green's function is obtained as,

$$G(\boldsymbol{x}-\boldsymbol{y})=\frac{1}{4\pi\sqrt{\det(\boldsymbol{K}_0)[(\boldsymbol{x}-\boldsymbol{y})^T\boldsymbol{K}_0^{-1}(\boldsymbol{x}-\boldsymbol{y})]}}. \tag{3}$$

(**See supplementary Note 1** for the details). For the isotropic materials ($K_{ij}=k_0\delta_{ij}$), the Greens' function reduces to a well-known function that is inversely proportional to the distance, $\frac{1}{4\pi k_0|x-y|}$. In the absence of interfacial thermal resistance, the Green's function is introduced to derive the Eshelby tensor $S_{ik}$ of an eigen-intensity problem that relates the intensity field $\boldsymbol{e} = -\nabla T$ and the eigen-intensity field $\boldsymbol{e}^*$ within the inclusion as[30,47],

$$S_{ik}(\boldsymbol{x}) = \frac{\partial}{\partial x_i}\int_V \frac{\partial G(\boldsymbol{x}-\boldsymbol{y})}{\partial y_j} d\boldsymbol{y}\, K_{0\,jk} \qquad \text{and} \qquad e_j = S_{ij} e_j^* \tag{4}$$

where $V$ is the volume of an inclusion whose thermal conductivity is identical to the matrix (see Figure 1(a)). The eigen-intensity field $\boldsymbol{e}^*$ can be considered as a fictitious temperature gradient field produced without external heat flux for an isolated inclusion, and the intensity field $\boldsymbol{e}$ is the temperature gradient within the inclusion when it is embedded in the matrix. With the Eshelby tensor, we can solve the eigen heat flux problem, which relates the heat flux within the inclusion $\boldsymbol{q}$ and eigen heat flux $\boldsymbol{q}^*$ as,

$$q_i = C_{ij} q_j^*. \tag{5}$$

where the $\boldsymbol{C}$ is known as the conjugate Eshelby tensor,

$$\boldsymbol{C} = -\boldsymbol{K}_0 \boldsymbol{S} \boldsymbol{K}_0^{-1} + \boldsymbol{I}. \tag{6}$$

By adopting the classical potential theory and the mathematical analogy with electrostatics[48] (See Appendix B in the reference), one can simplify Eq. (4) for an ellipsoidal inclusion as

$$\boldsymbol{S} = \frac{\det(\boldsymbol{a})}{2}\int_0^\infty \frac{(\boldsymbol{a}^2 + s\boldsymbol{K}_0)^{-1}\boldsymbol{K}_0}{\sqrt{\det(\boldsymbol{a}^2 + s\boldsymbol{K}_0)}} ds. \tag{7}$$

The tensor $\boldsymbol{a} = \begin{bmatrix} a_1 & 0 & 0 \\ 0 & a_2 & 0 \\ 0 & 0 & a_3 \end{bmatrix}$ is a diagonal matrix with its diagonal elements being the half the length of the principal axes of an ellipsoidal inclusion whose volume is defined as $\left\{(x_1, x_2, x_3) : \sum_{i=1}^{3} \frac{x_i^2}{a_i^2} \leq 1\right\}$. For example, for the spherical inclusion, $\boldsymbol{a}$ can be expressed as $a_{ij} = R\delta_{ij}$, where $R$ is the radius of inclusion.

In this study, we consider the Eshelby tensor for a spherical inclusion in the orthotropic matrix with three independent thermal conductivity coefficients. The thermal conductivity tensors of matrix ($\boldsymbol{K}_0$) and inhomogeneity ($\boldsymbol{K}_1$) are defined as follows,

$$\boldsymbol{K}_0 = \begin{bmatrix} k_1 & 0 & 0 \\ 0 & k_2 & 0 \\ 0 & 0 & k_3 \end{bmatrix} \quad \text{and} \quad \boldsymbol{K}_1 = \begin{bmatrix} \kappa_1 & \kappa_{12} & \kappa_{13} \\ \kappa_{12} & \kappa_2 & \kappa_{23} \\ \kappa_{13} & \kappa_{23} & \kappa_3 \end{bmatrix}. \tag{8}$$

Now, we derive a closed form expression of $S_{ij}$ for a spherical inclusion by plugging the thermal conductivity tensor $\boldsymbol{K}_0$ into the Eq. (7) with $a_{ij} = R\delta_{ij}$ as,

$$S_{IJ} = \delta_{IJ}\frac{R^3}{2}\int_0^\infty \frac{(R^2 + sk_I)^{-1}k_I}{\sqrt{\prod_{\ell=1}^3 (R^2 + sk_\ell)}}ds = \delta_{IJ}\frac{(k_I)^{3/2}}{2\sqrt{\det(\boldsymbol{K}_0)}}\int_0^\infty \frac{1}{(s'+1)\sqrt{\prod_{\ell=1}^3\left(s' + \frac{k_I}{k_\ell}\right)}}ds' \tag{9}$$

where for the last equality, we set $s' = sk_I/R^2$. Because both $k_I$ and $R^2 > 0$, range of integration does not change (**See supplementary Note 2** for the details). The repeated capital indices $I, J$ are not summed over. We note that $\boldsymbol{S}$ is a diagonal matrix because $\boldsymbol{K}_0$ is a diagonal matrix. By defining the anisotropy factors of the matrix as $A = k_I/k_L, B = k_I/k_M$ with $I \neq L, I \neq M,$ and $L \neq M$, we can simplify the Eshelby tensor as

$$S_{IJ} = \delta_{IJ}\frac{(k_I)^{\frac{3}{2}}}{\sqrt{\det(\boldsymbol{K}_0)}}\frac{1}{A-B}\left(\frac{\mathrm{E}\left(\cos^{-1}(\sqrt{B}), \frac{A-1}{B-1}\right)}{\sqrt{1-B}} - \frac{\mathrm{E}\left(\cos^{-1}(\sqrt{A}), \frac{B-1}{A-1}\right)}{\sqrt{1-A}}\right) \tag{10}$$

where $E(\theta, m) = \int_0^\theta \sqrt{1 - m\sin^2\theta'}\,d\theta'$ is elliptic integral of the 2nd kind (See **supplementary Note 3** for the details). For example, when $\mathrm{I} = \mathrm{J} = 2$, either $\mathrm{L} = 1, \mathrm{M} = 3$ or $\mathrm{L} = 3, \mathrm{M} = 1$ is used for calculating the $S_{22}$ component, and either $\mathrm{L}, \mathrm{M}$ combination results in the same result. We note that, at the limit $B \to A$ (or $A \to B$), the Eshelby tensor reduces to the transversely isotropic matrix result in a closed form solution (See **supplementary Note 3**). At the limit $A, B \to 1$, the Eshelby tensor converges to an isotropic matrix case such that $S_{ij} = \frac{1}{3}\delta_{ij}$[22,30], (see **Supplementary Fig. S1**). The three independent values ($S_{11}$, $S_{22}$, $S_{33}$) are plotted in terms of $k_1/k_2$ and $k_1/k_3$, and we validate our analytical solutions against the numerical evaluation of Eq. (4) (see **Supplementary Fig. S1**).

It has been proven that the heat flux $\boldsymbol{q}^{\mathrm{inh}}$ within an ellipsoidal inhomogeneity embedded in an infinite matrix under the presence of a constant external far field heat flux $\boldsymbol{q}^{\mathrm{ext}}$ is uniform, and so is the Eshelby tensor $\boldsymbol{S}$, regardless of the materials symmetry of the matrix[29,30] (see Figure 1(b)). Heat fluxes, $\boldsymbol{q}^{\mathrm{inh}}$ and $\boldsymbol{q}^{\mathrm{ext}}$, are related by the localization tensor $\boldsymbol{B}$[22,30] as $\boldsymbol{q}^{\mathrm{inh}} = \boldsymbol{B}\boldsymbol{q}^{\mathrm{ext}}$, where $\boldsymbol{B} = [\boldsymbol{I} + \boldsymbol{C}\boldsymbol{K}_0(\boldsymbol{K}_1^{-1} - \boldsymbol{K}_0^{-1})]^{-1}$. Using the Eshelby tensor in the orthotropic matrix in Eq. (10), we predict the heat flux within the inhomogeneity with any arbitrary thermal conductivity tensor $\boldsymbol{K}_1$. For an inhomogeneity with isotropic or cubic symmetry whose thermal conductivity tensor is given as $K_{1_{ij}} = \kappa\delta_{ij}$, the heat flux expression can be simplified as

$$q_I^{\mathrm{inh}} = \frac{\kappa}{k_I + S_{II}(\kappa - k_I)} q_I^{\mathrm{ext}} \tag{11}$$

Our solution is validated against the numerical calculations based on FEA where we consider a single inhomogeneity surrounded by an orthotropic medium in a cubic shape. We set the edge of the medium to be 15 times larger than the diameter of the inhomogeneity to reasonably describe the infinite medium[22] as depicted in Figure 2. The FEA is performed by COMSOL software with a total of 416,606 tetrahedral quadratic elements in the matrix and inhomogeneity. In this simulation, the unit heat flux boundary conditions are considered in the $x$, $y$, and $z$ directions to study the effect of the anisotropy ($q_I^{\mathrm{ext}} = 1$ W/m$^2$, $q_J^{\mathrm{ext}} = 0$ with $I \neq J$). We carry out the calculations using $k_1 = 1, k_2 = 2, k_3 = 3$ and $\kappa = 10$ (W/mK). As expected, the calculated heat flux within the inhomogeneity is uniform and dependent on the external heat flux direction (see Figure 2(b)), and matches very well with the theoretical prediction (see Figure 3).

The effective thermal conductivity of a composite with multiple inhomogeneities can be predicted by considering the interaction between the inhomogeneities. In a mean field approach, as in the Mori-Tanaka method, the heat flux within the inhomogeneity is related to the average heat flux within the matrix. The Mori-Tanaka model is known to predict effective properties well at a relatively low inhomogeneity volume concentration (< 20%) and is more convenient than the self-consistent method, which relies on a nonlinear implicit equation. In the absence of the interfacial resistance, the effective thermal conductivity based on the Mori-Tanaka method can be obtained as[19,30]

$$K_I^{\mathrm{eff}} = \frac{k_I[(1 - S_{II})c_0 k_I + (c_1 + S_{II}c_0)\kappa]}{[k_I + S_{II}c_0(\kappa - k_I)]} \tag{12}$$

Here, $c_0$ and $c_1$ refer to volume concentrations of the matrix and inhomogeneity, respectively; thus, $c_0 + c_1 = 1$.

## Effective Thermal Conductivity in the Presence of Interfacial Resistance

We now turn our attention to the realistic system, where interfacial thermal resistance is present[31,32]. The interfacial thermal resistance $\alpha$ is defined as

$$T^{\mathrm{out}} - T^{\mathrm{in}} = -\alpha \boldsymbol{q} \cdot \boldsymbol{n} \tag{13}$$

where $T^{\mathrm{out}}$ and $T^{\mathrm{in}}$ refer to temperatures at the outer and inner surfaces of the interface, respectively, $\boldsymbol{q}$ is the heat flux at the interface, and the $\boldsymbol{n}$ is the outward surface normal vector (see Figure 1(c) and (d)). The SI unit of interfacial thermal resistance $\alpha$ is [$m^2K/W$].

The interfacial resistance augments an additional surface integration term in the eigen-intensity problem, as follows,

$$e_m(\boldsymbol{x}) = \frac{\partial}{\partial x_m}\int_V \frac{\partial G(\boldsymbol{x}-\boldsymbol{y})}{\partial y_i} d\boldsymbol{y} K_{0_{ij}} e_j^* + \frac{\partial}{\partial x_m} \alpha K_{0_{ij}} K_{0_{sr}} \int_{\partial V} \frac{\partial G(\boldsymbol{x}-\boldsymbol{y})}{\partial y_j} n_i(\boldsymbol{y}) n_s(\boldsymbol{y}) \big(e_r(\boldsymbol{y}) - e_r^*(\boldsymbol{y})\big) d\boldsymbol{y}. \quad (14)$$

It has been found that the heat flux within a spherical inclusion is uniform in the presence of interfacial thermal resistance[22]. Although a previous study[22] claims that the heat flux within an elliptical inclusion is also uniform in the presence of interfacial thermal resistance, our numerical tests reveal that it is non-uniform (See **Supplementary Note 4**). Because the intensity field in the spherical inclusion is uniform, we can simplify Eq. (14) as follows:

$$e_m(\boldsymbol{x}) = \frac{\partial}{\partial x_m}\int_V \frac{\partial G(\boldsymbol{y}-\boldsymbol{x})}{\partial y_i} \mathrm{d}\boldsymbol{y} K_{0_{ij}} e_j^* + \alpha K_{0_{ij}} K_{0_{sr}} \frac{\partial}{\partial x_m} \int_{\partial V} \frac{\partial G(\boldsymbol{y}-\boldsymbol{x})}{\partial y_j} n_i(\boldsymbol{y}) n_s(\boldsymbol{y}) \mathrm{d}\boldsymbol{y}\, (e_r - e_r^*)$$

$$= S_{mj} e_j^* + \alpha K_{0_{ij}} K_{0_{sr}} M_{ijms} (e_r - e_r^*) \quad (15)$$

$$\text{where } M_{ijms} \equiv \frac{\partial}{\partial x_m}\int_{\partial V} \frac{\partial G(\boldsymbol{y}-\boldsymbol{x})}{\partial y_j} n_i(\boldsymbol{y}) n_s(\boldsymbol{y}) \mathrm{d}\boldsymbol{y}$$

From $e_m = S_{mj}^M e_j^*$, the modified Eshelby tensor is given as

$$\boldsymbol{S}^M = (\boldsymbol{I} - \alpha \boldsymbol{K}_0 \colon \boldsymbol{M} \boldsymbol{K}_0)^{-1} (\boldsymbol{S} - \alpha \boldsymbol{K}_0 \colon \boldsymbol{M} \boldsymbol{K}_0). \quad (16)$$

Where the colon in Eq. (16) is a double dot product. We also find that the Eq. (16) can be further simplified (See **Supplementary Note 5** for details), as follows:

$$\boldsymbol{S}^M = \left(\boldsymbol{I} + \frac{\alpha}{R}(\boldsymbol{I} - \boldsymbol{S})\boldsymbol{K}_0\right)^{-1} \left(\boldsymbol{S} + \frac{\alpha}{R}(\boldsymbol{I} - \boldsymbol{S})\boldsymbol{K}_0\right). \quad (17)$$

Unlike the single inhomogeneity problem without interfacial thermal resistance, the modified localization tensor $\boldsymbol{B}^M$ should be obtained by decomposing the original problem into three independent problems[22]. We obtain the relationship between the heat flux $\boldsymbol{q}^{\mathrm{inh}}$ within a spherical inhomogeneity and the far field heat flux $\boldsymbol{q}^{\mathrm{ext}}$ as $\boldsymbol{q}^{\mathrm{inh}} = B^M \boldsymbol{q}^{\mathrm{ext}}$, where $\boldsymbol{B}^M = [\boldsymbol{C}(\boldsymbol{C}^{\boldsymbol{M}})^{-1} + \boldsymbol{C}\boldsymbol{K}_0(\boldsymbol{K}_1^{-1} - \boldsymbol{K}_0^{-1})]^{-1}$. $\boldsymbol{C}^{\boldsymbol{M}}$ is the modified conjugate Eshelby tensor which is given as $\boldsymbol{C}^{\boldsymbol{M}} = -\boldsymbol{K}_0 \boldsymbol{S}^{\boldsymbol{M}} \boldsymbol{K}_0^{-1} + \boldsymbol{I}$. After substituting the conjugate Eshelby tensor and the modified conjugate Eshelby tensor into the modified localization tensor equation, the heat flux within the inhomogeneity in the presence of interfacial thermal resistance can be obtained as follows,

$$q_I^{\text{inh}} = \frac{\kappa}{S_{II}\kappa + k_I(1 - S_{II})(1 + \alpha\kappa/R)} q_I^{\text{ext}} \tag{18}$$

The theoretical predictions of heat fluxes along three directions match very well with the FEA calculation results with the same boundary conditions for a wide range of interfacial thermal resistance $\alpha$, as shown in Figure 4. At the limit of zero resistance, i.e., $\alpha \to 0$, the heat flux within the inhomogeneity converges to the zero interfacial resistance case depicted in Figure 3. At the opposite limit, as $\alpha \to \infty$, the heat flux within the inhomogeneity reduces to zero, which implies that the infinite interfacial thermal resistance is equivalent to a void or a perfect thermal insulator as an inhomogeneity. Hence, the effective conductivity approaches that of a porous medium.

We then derive a closed form solution for the effective thermal conductivity based on a mean field micromechanics model, the Mori-Tanaka method. Following the previous study[22], the effective thermal conductivity of a composite with interfacial thermal resistance can be determined by

$$\boldsymbol{K}_{\text{eff}} = \left\{\boldsymbol{K}_1^{-1} + \frac{\alpha}{R}\boldsymbol{I} - c_0\left[\boldsymbol{K}_1^{-1} - \boldsymbol{K}_0^{-1} + \frac{\alpha}{R}\boldsymbol{I}\right]\boldsymbol{L}\right\}^{-1} \tag{19}$$

where $\boldsymbol{L} = \{c_1[\boldsymbol{C}(\boldsymbol{C}^{\boldsymbol{M}})^{-1} + \boldsymbol{C}\boldsymbol{K}_0(\boldsymbol{K}_1^{-1} - \boldsymbol{K}_0^{-1})]^{-1} + c_0\boldsymbol{I}\}^{-1}$ and $R$ is the radius of the inhomogeneity. The effective thermal conductivity of the composite is finally obtained in the closed form as

$$K_I^{\text{eff}} = \frac{k_I\left[\kappa + (1 - S_{II})c_0\left((1 + \alpha\kappa/R)k_I - \kappa\right)\right]}{\left[k_I(1 + \alpha\kappa/R) - S_{II}c_0\left((1 + \alpha\kappa/R)k_I - \kappa\right)\right]} \tag{20}$$

Although we consider a thermally isotropic inhomogeneity in the final solution for the sake of simplicity, the effective thermal conductivity with anisotropic inhomogeneity can be easily predicted by using an anisotropic $\boldsymbol{K}_1$.

We plot the effective thermal conductivities, $K_1^{\text{eff}}$, $K_2^{\text{eff}}$ and $K_3^{\text{eff}}$, as a function of the inhomogeneity's volume fraction $c_1$ and the interfacial thermal resistance $\alpha$, as shown in Figure 5. At the limit of zero interfacial resistance, i.e., $\alpha \to 0$, Eq. (20) becomes identical to Eq. (12), where we assumed no interfacial resistance. At the opposite limit of $\alpha \to \infty$, because no heat flux is permitted within the inhomogeneity, the effective conductivity tensor converges to that of a porous medium,

$$\lim_{\alpha\to\infty} \frac{k_I\big[\kappa + (1 - S_{II})c_0\big((1 + \alpha\kappa/R)k_I - \kappa\big)\big]}{\big[k_I(1 + \alpha\kappa/R) - S_{II}c_0\big((1 + \alpha\kappa/R)k_I - \kappa\big)\big]} = \frac{(1 - S_{II})c_0 k_I}{1 - S_{II}c_0} \tag{21}$$

Here, the right-hand side is identical with the zero interfacial resistance solution, Eq. (12), with $\kappa = 0$ (i.e., porous medium solution)

$$\left.\frac{k_I[\kappa + (1 - S_{II})c_0(k_I - \kappa)]}{[k_I - S_{II}c_0(k_I - \kappa)]}\right|_{\kappa=0} = \frac{(1 - S_{II})c_0 k_I}{1 - S_{II}c_0} \tag{22}$$

These two limiting cases of $\alpha \to 0$ and $\alpha \to \infty$ define the upper bound and the lower bound of the thermal conductivity values, respectively. In plotting Figure 5, we set $k_1 = 1, k_2 = 2, k_3 = 3$ (W/mK), $\kappa = 10$ (W/mK), and $R = 1\ mm$. Because the inhomogeneity is more conductive than the matrix ($\kappa > k_1, k_2, k_3$), the effective thermal conductivity increases with the volume fraction in the range where the interfacial thermal resistance is low. However, at high enough interfacial thermal resistance, the effective thermal conductivity decreases with the volume fraction because the heat flux through the inhomogeneity is significantly limited. We calculate the critical interfacial thermal resistance that makes the effective thermal conductivity of the composite $K_I^{eff}$ identical to the thermal conductivity of the matrix $k_I$, as follows:

$$\alpha^{critical} = R\left(\frac{1}{\kappa} - \frac{1}{k_I}\right) \tag{23}$$

When the interfacial thermal resistance is higher than the critical value, the effective thermal conductivity of the composite decreases as the volume fraction increases, even though the inhomogeneity is thermally more conductive than the matrix. We note that the critical interfacial resistance decreases for smaller particles, because the interface area increases with the decreasing size of the inhomogeneity for a given volume fraction.

We validate our analytical solution presented in Eq. (20) by comparing it with the effective thermal conductivity calculated numerically by FEA, as depicted in Figure 6. We obtain the effective conductivity by averaging the results from 10 independent RVEs, each containing multiple spherical inhomogeneities that are randomly distributed within a cube (see Figure 7). We assign a uniform temperature boundary condition at two parallel surfaces while applying a periodic boundary conditions along the other two directions. We then compute the heat flux to obtain the effective thermal conductivity of each RVE. As shown in Figure 6(a), (b) and (c), the FEA results match very well with our solution for up to 10% of the inhomogeneity volume fraction for a wide range of interfacial thermal resistances. We also investigate the effect of the inhomogeneity's size at a fixed volume fraction of 5% and an interfacial thermal resistance ($10^{-3}$ $m^2K/W$), as depicted in Figure 6(d), (e) and (f). When the interfacial resistance is absent, both theoretical predictions and numerical results find that the effective conductivity is independent of the inhomogeneity's size. In contrast, in the presence of interfacial resistance, the effective thermal conductivity decreases as the radius of inhomogeneity decreases, because the interface fraction is bigger for smaller inhomogeneities for a fixed volume.

## Conclusion

In conclusion, we have investigated the heat conduction problem of composites with orthotropic matrices and a spherical inhomogeneities in the presence of interfacial thermal resistance. We derive the modified Eshelby tensor of the eigen-intensity problem as well as the effective thermal conductivity based on a micromechanics approach by considering the interfacial thermal resistance effect, and validate our solution against FEA calculation results. We also demonstrate that the effective conductivity solution has the correct limiting behaviour at both the zero and infinite interfacial thermal resistance limit. The solution in the present paper is applicable to the composites with either transversely isotropic or isotropic matrices and an inhomogeneity with an arbitrary thermal conductivity tensor. We plan to extend the present study by considering the size effects of nanoscale inclusions for nanocomposites[49] in obtaining an analytic solution and by coupling molecular dynamics simulations of ceramic composites or polymer composites with the analytic solution. We believe that our study can provide an effective and accurate way of predicting the thermal conduction of composites, and it can be applied to better design technologically important materials such as polymer-based composite and thermoelectric materials.

**Figures and captions**

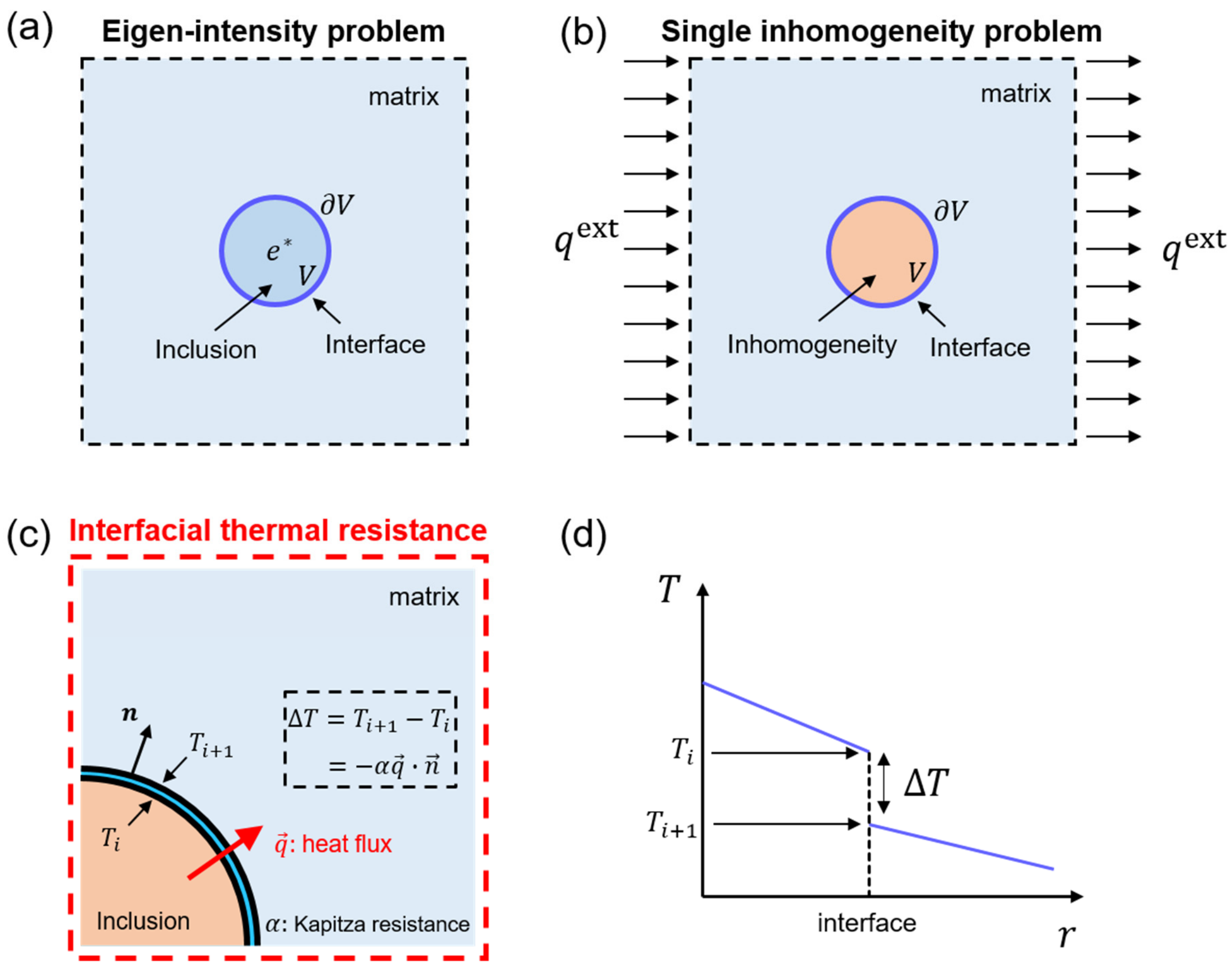

Figure 1. Schematic of (a) an eigen-intensity problem, (b) a single inhomogeneity problem and (c) interfacial thermal resistance. (d) The temperature jump at the interface.

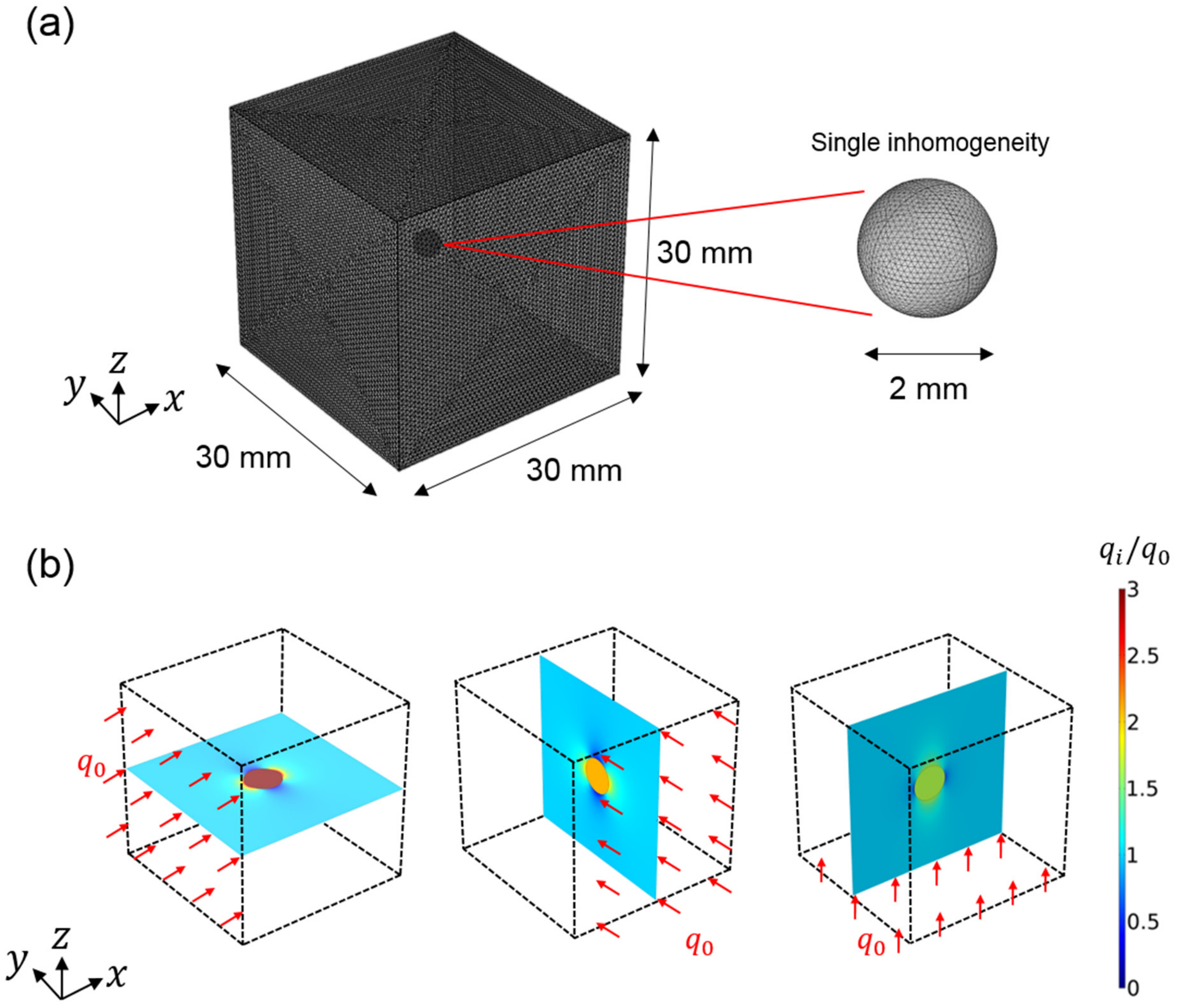

Figure 2. (a) The mesh configuration for a single inhomogeneity and the matrix used in FEA. (b) Heat flux distribution within the matrix and the inhomogeneity at $x, y, z$ plane for three different heat flux directions. The thermal conductivities used for the results are $k_1 = 1, k_2 = 2, k_3 = 3$ and $\kappa = 10$ (W/mK).

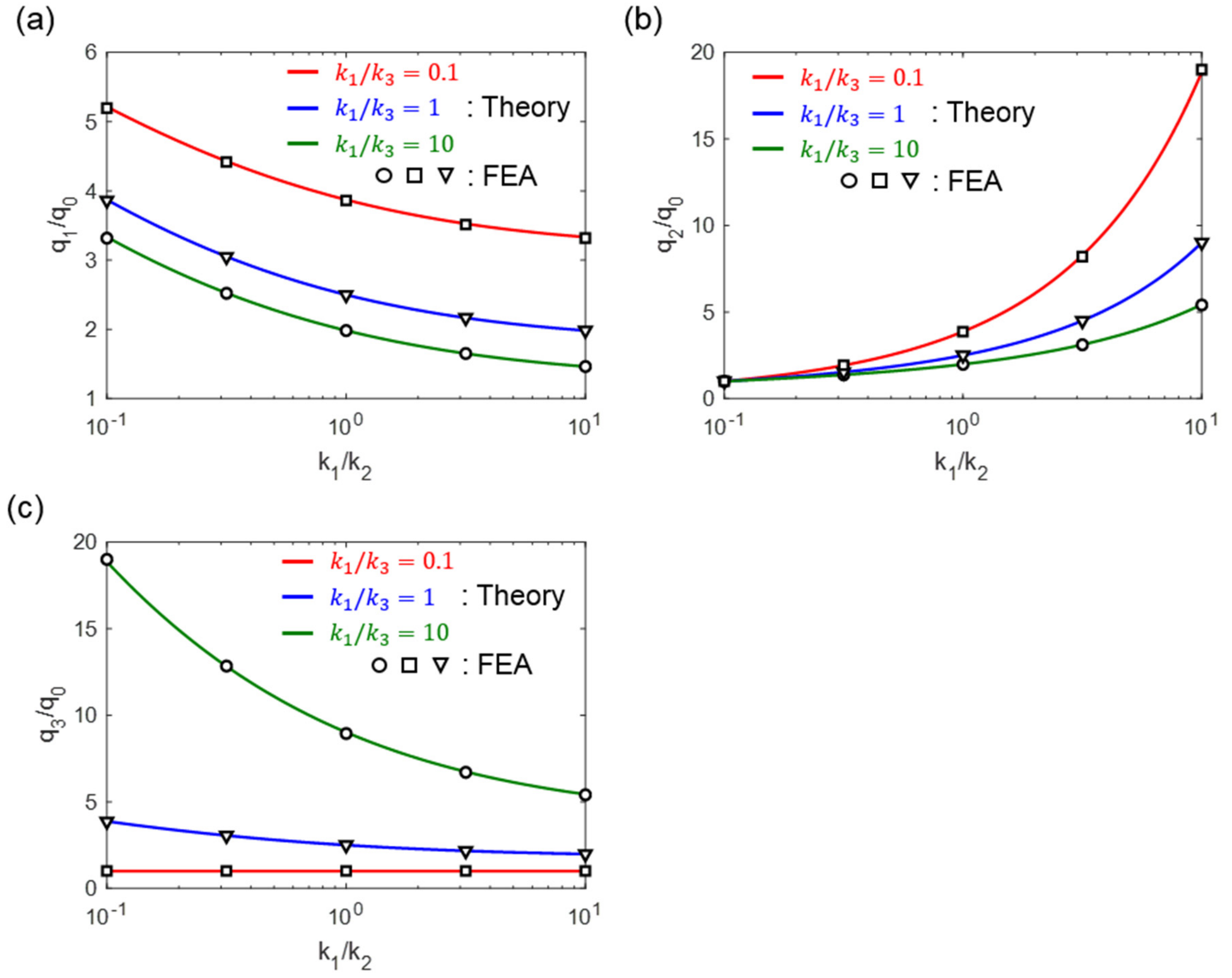

Figure 3. Normalized uniform heat flux values ((a) $q_1/q_0$, (b) $q_2/q_0$, (c) $q_3/q_0$) within the inhomogeneity as a function of anisotropy factors where $q_0$ is the magnitude of the heat flux at the boundary. The isotropic thermal conductivity of a single inhomogeneity is 10 (W/mK).

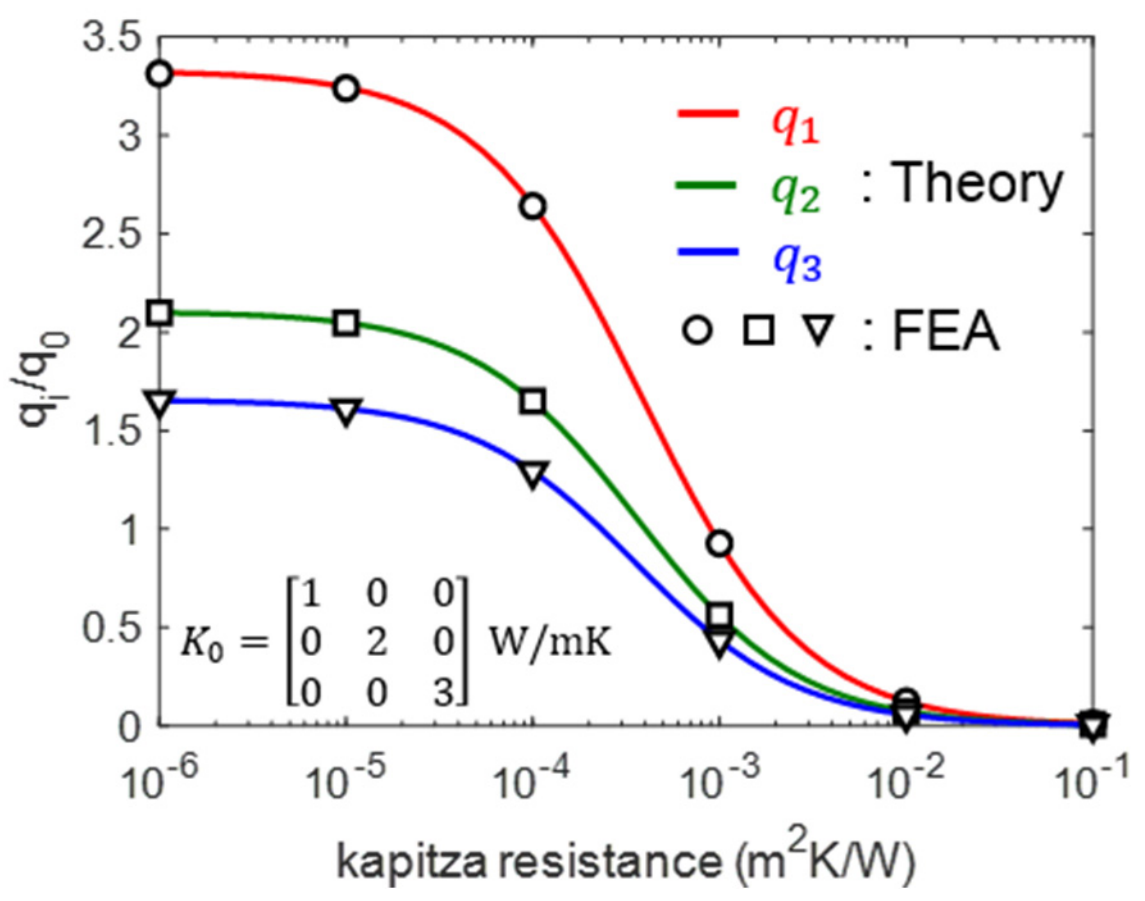

Figure 4. Normalized heat flux within the inhomogeneity with respect to the interfacial thermal resistance in a single inhomogeneity problem. The isotropic thermal conductivity of the inhomogeneity is 10 (W/mK).

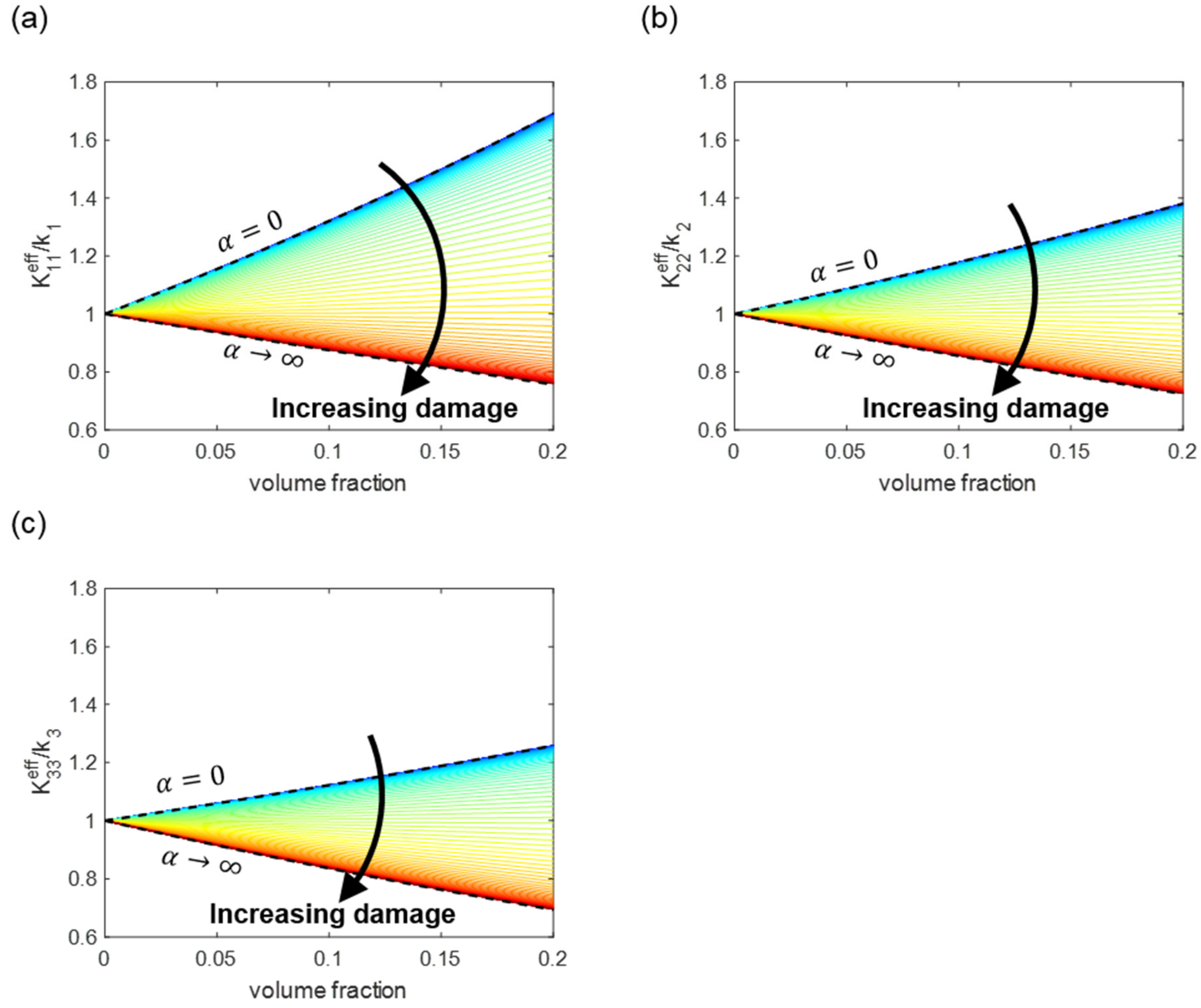

Figure 5. Effective thermal conductivity of composite ((a) $K_{11}^{\text{eff}}$, (b) $K_{22}^{\text{eff}}$, (c) $K_{33}^{\text{eff}}$) having interfacial thermal resistance as a function of volume fraction of inhomogeneity. The thermal conductivities of the matrix and inhomogeneity are $k_1 = 1, k_2 = 2, k_3 = 3$ and $\kappa = 10$ (W/mK), and the radius of the particle is 1 (mm).

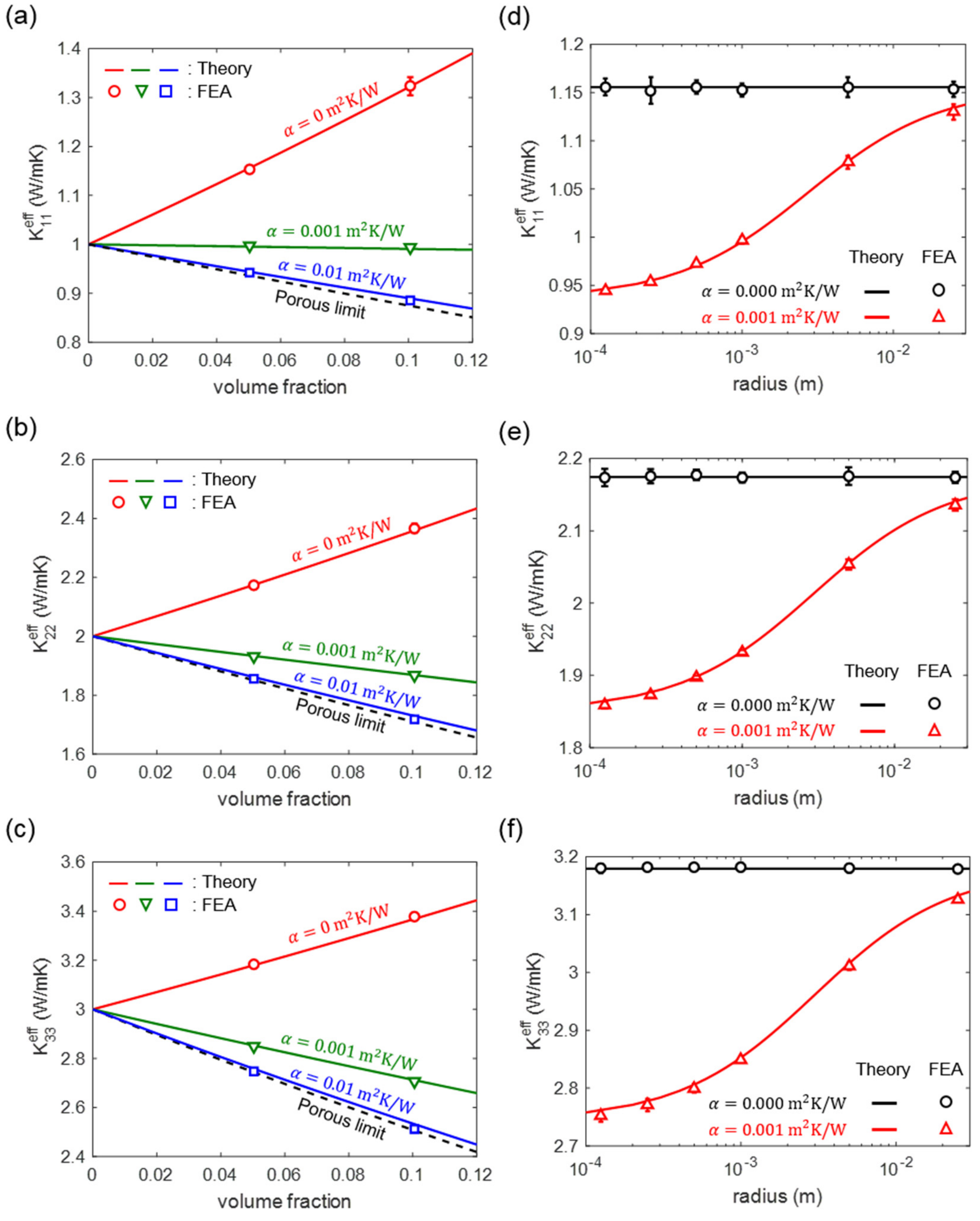

Figure 6. Effective thermal conductivity ((a)$K_1^{\text{eff}}$, (b)$K_2^{\text{eff}}$, (c)$K_3^{\text{eff}}$) of particle reinforced composite for different interfacial thermal resistances. The radius of the particle used in (a), (b) and (c) is 1 (mm). The effective thermal conductivity ((d)$K_1^{\text{eff}}$, (e)$K_2^{\text{eff}}$, (f)$K_3^{\text{eff}}$) as a function of radius of the particle under fixed volume fraction of 5%. The thermal conductivities of the orthotropic matrix and inclusion are $k_1 = 1, k_2 = 2, k_3 = 3$ and $\kappa = 10$ (W/mK) respectively.

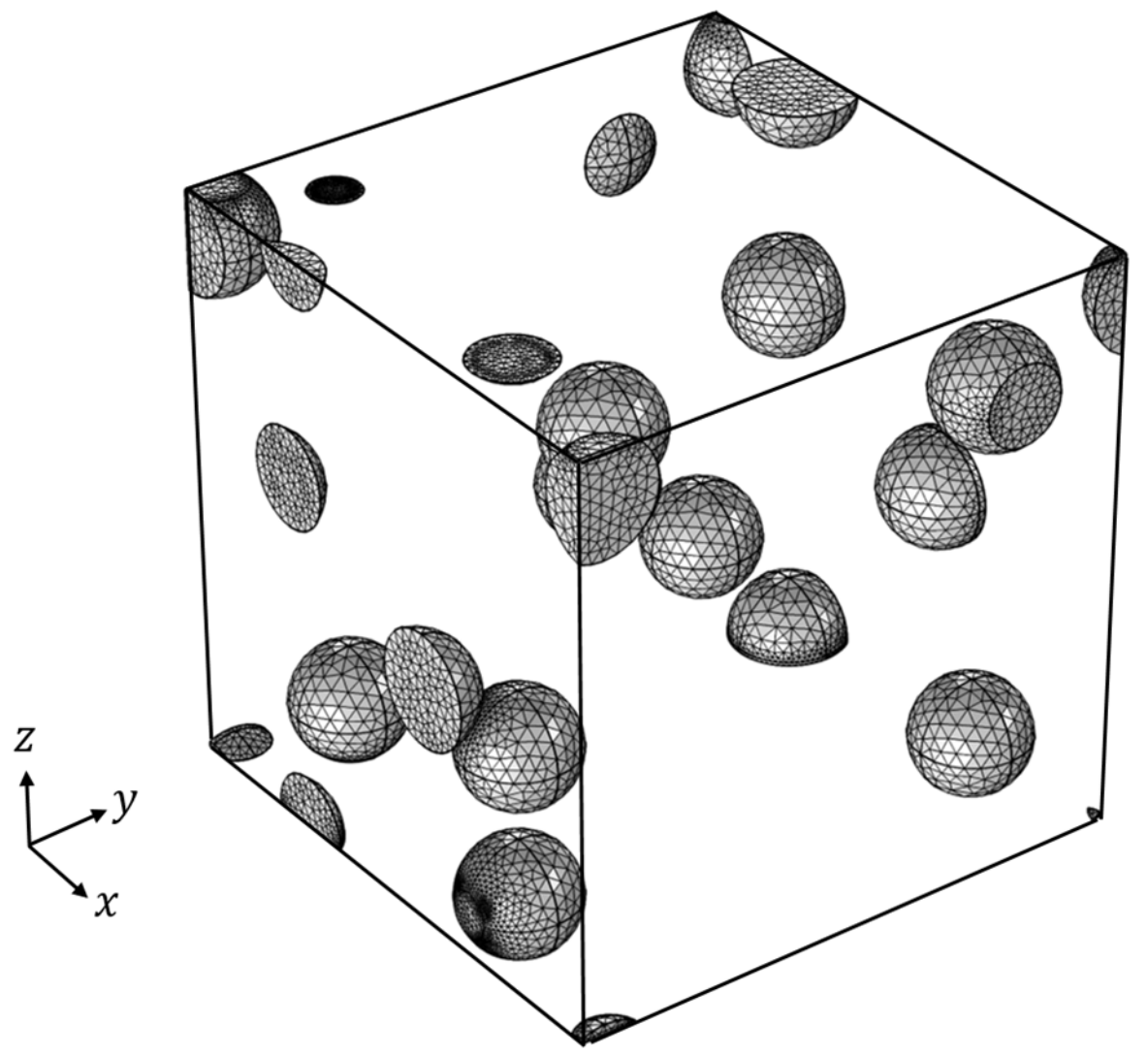

Figure 7. Mesh configuration of inhomogeneities in representative volume element for FEA. The volume fraction is 5% and the particle radius is 1 (mm).

**Acknowledgements**

This work is supported by the National Research Foundation of Korea (NRF) funded by the Ministry of Science and ICT (2016M3D1A1900038 and 2016R1C1B2011979).

**Author contributions Statement**

S.L and S.R designed the research, interpret the results, and wrote the manuscript. S.L. and J.L. carried out the analytic derivation. All authors discussed and analysed the results.

**Additional information**

The authors declare no competing interests.

**Supplementary information**

# A micromechanics-based analytical solution for the effective thermal conductivity of composites with orthotropic matrices and interfacial thermal resistance

Sangryun Lee[1,†], Jinyeop Lee[2,†], Byungki Ryu[3] and Seunghwa Ryu[1,*]

**Affiliations**

[1]Department of Mechanical Engineering and [2]Department of Mathematical Sciences, Korea Advanced Institute of Science and Technology (KAIST), 291 Daehak-ro, Yuseong-gu, Daejeon 34141, Republic of Korea

[3]Thermoelectric Conversion Research Center, Korea Electrotechnology Research Institute (KERI), Changwon-si, Gyoengsangnam-do, 51543, Republic of Korea

[†]These authors contributed equally to this work.

[*]Corresponding author e-mail: ryush@kaist.ac.kr

**Supplementary Table 1: Mathematical analogy between steady-state heat conduction, elastostatics and electrostatics.**

| **Heat conduction** | **Elastostatics** | **Electrostatics** |
|---|---|---|
| Intensity field<br>$e_i$ $\left(e_i = -\frac{\partial T}{\partial x_i}\right)$ | Elastic strain<br>$\varepsilon_{ij}$ $\left(\varepsilon_{ij} = \frac{1}{2}\left(\frac{\partial u_j}{\partial x_i} + \frac{\partial u_i}{\partial x_j}\right)\right)$ | Electric field<br>$E_i$ $\left(E_i = -\frac{\partial V}{\partial x_i}\right)$ |
| Temperature field<br>$T$ | Elastic displacement<br>$u_i$ | Electric potential<br>$V$ |
| Heat flux<br>$q_i$ | Stress tensor<br>$\sigma_{ij}$ | Electric displacement<br>$D_i$ |
| Fourier equation<br>$q_{i,i} = 0$ | Equilibrium equation<br>$\sigma_{ij,i} = 0$ | Maxwell equation<br>$D_{i,i} = 0$ |
| Thermal conductivity tensor<br>$K_{ij}$ | Stiffness tensor<br>$L_{ijkl}$ | Permittivity tensor<br>$\epsilon_{ij}$ |
| Constitutive equation<br>$q_i = K_{ij}e_j$ | Constitutive equation<br>$\sigma_{ij} = L_{ijkl}\varepsilon_{kl}$ | Constitutive equation<br>$D_i = \epsilon_{ij}E_j$ |

**Supplementary Figure 1: Eshelby tensor for anisotropic matrix**

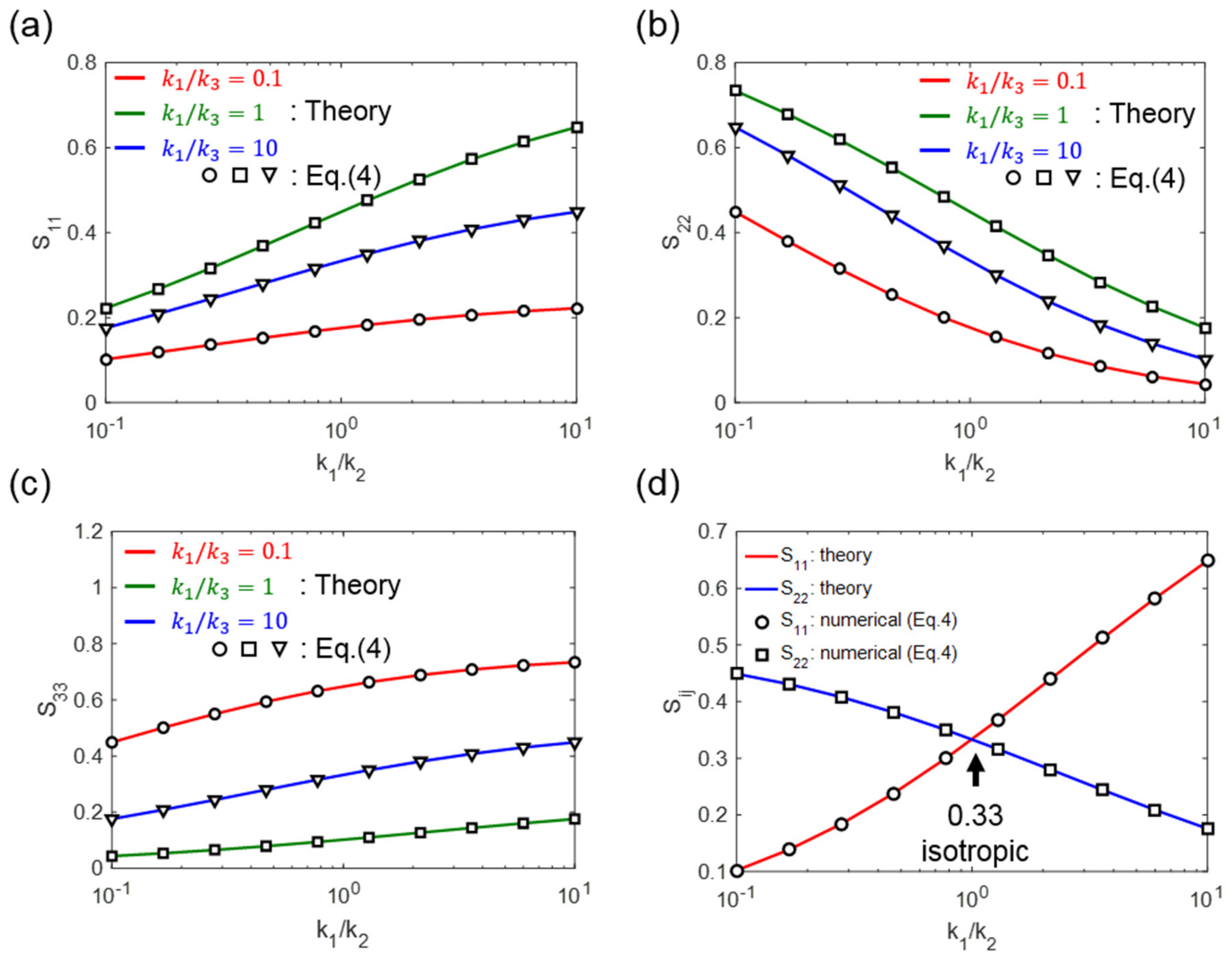

**Supplementary Figure S1.** The Three independent Eshelby tensor components ((a) $S_{11}$, (b) $S_{22}$, (c) $S_{33}$) for a spherical inclusion in an orthotropic matrix. (d) Two independent components ($S_{11}$, $S_{22}$) in the Eshelby tensor for a transversely isotropic matrix with a spherical inclusion as a function of $k_1/k_2$.

**Supplementary Figure 2: Heat flux within ellipsoidal inclusion having interfacial thermal resistance**

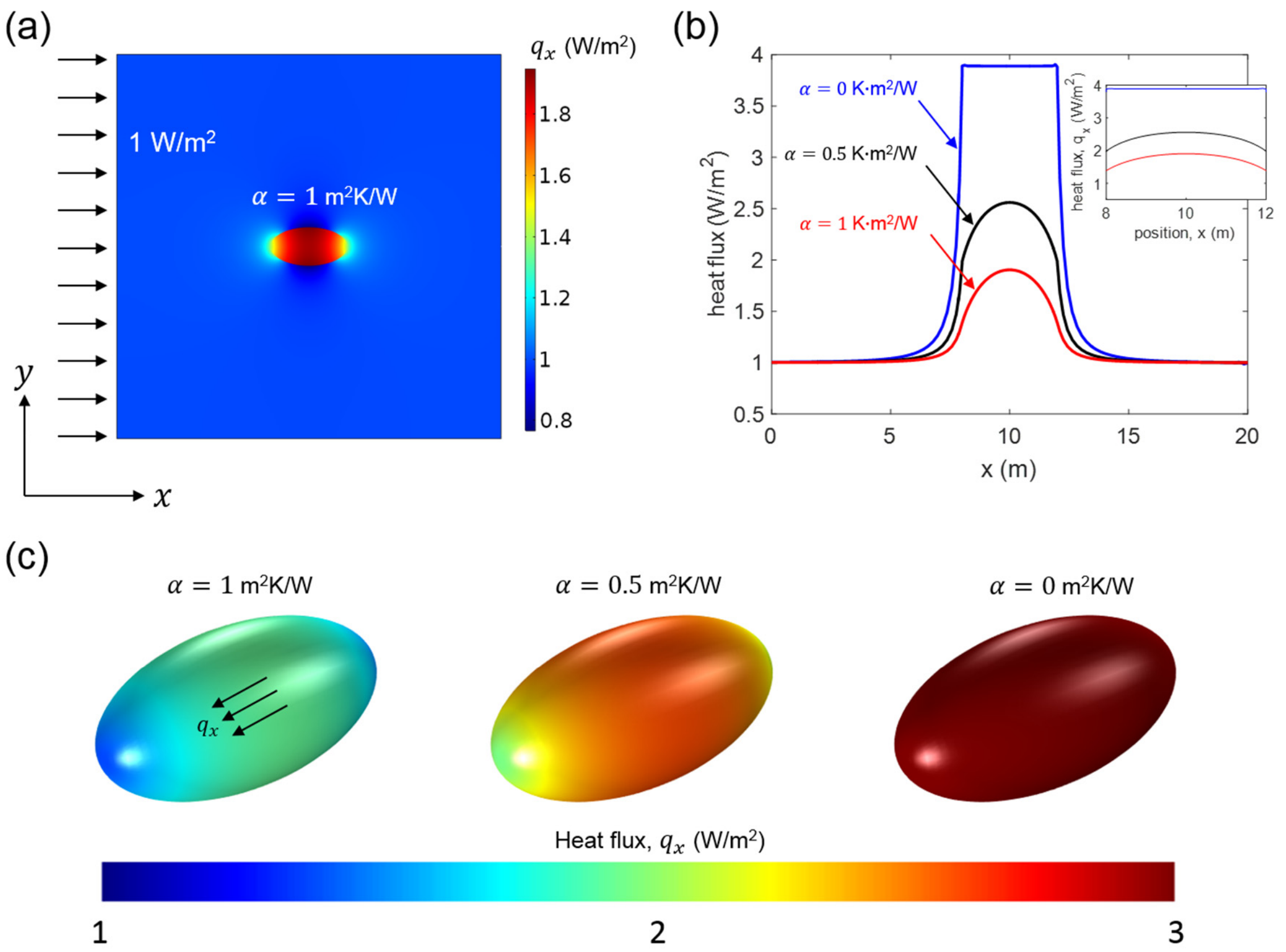

**Supplementary Figure S2.** (a) Heat flux in the x direction at the z=0 plane. (b) Heat flux along the long axis of inclusion (aspect ratio 2) with various interfacial resistances. (c) Heat flux at the surface of the inclusion with different Kapitza resistances. The isotropic thermal conductivities of the matrix and inclusion are 1 and 10 (W/mK) respectively, and the heat flux at the boundary is 1 W/m$^2$.

**Supplementary Figure 3: $M_{1111}$ component for ellipsoidal inclusion under uniform heat flux assumption**

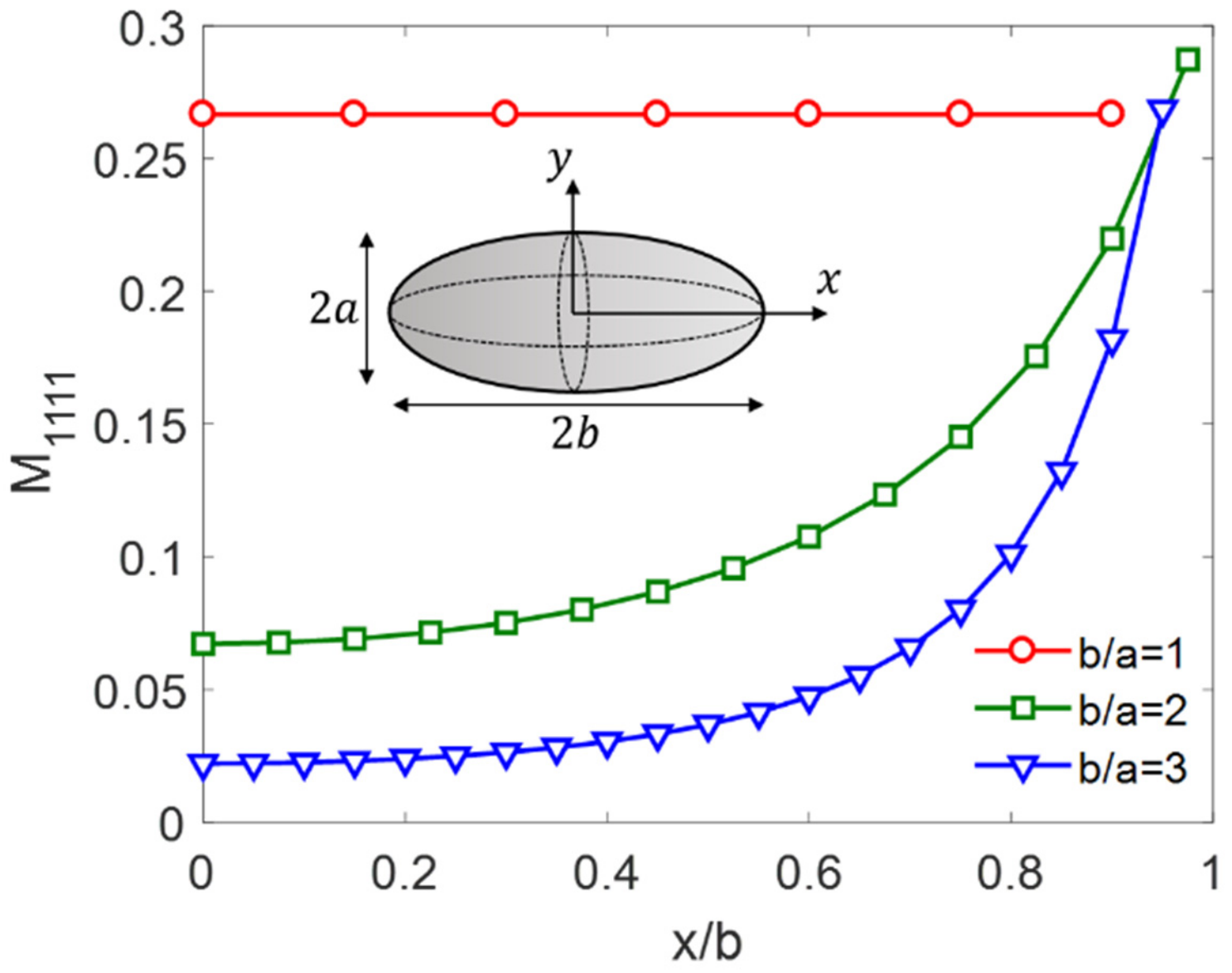

**Supplementary Figure S3.** $M_{1111}$ for ellipsoids with different aspect ratios. The isotropic thermal conductivity of the inclusion is 1 W/mK.

**Supplementary Note 1: Green's function for anisotropic matrix.**

Comparing to Eq. (2) of previous research[1],

$$\epsilon_{kl}\frac{\partial}{\partial x_k}\frac{\partial}{\partial x_l}V(\vec{r}) = -Q(\vec{r}),$$

We use the mathematical analogy between electrostatics and heat conduction so we can formally set $k = i$, $\ell = j$, $\overleftrightarrow{\epsilon} = \boldsymbol{K}_0$, $V = G$, and $Q = 1$. Then as Eq. (3) of the previous research

$$V(\vec{r}) = \frac{1}{4\pi\sqrt{\det(\overleftrightarrow{\epsilon})\,[\vec{r}^T\overleftrightarrow{\epsilon}^{-1}\vec{r}]}}$$

obtained from Eq. (2) of the previous research, we get

$$G(\vec{r}) = \frac{1}{4\pi\sqrt{\det(K_0)\,[\vec{r}^T K_o^{-1}\vec{r}]}}$$

from

$$K_{0ij}\frac{\partial}{\partial x_i}\frac{\partial}{\partial x_j}G(\vec{r}) + \delta(\vec{r}) = 0.$$

By putting $\vec{r} = \boldsymbol{x} - \boldsymbol{y}$, we get Eq. (3) from Eq. (2).

The detailed proof is in Appendix A of the previous research[1].

**Supplementary Note 2: Eshelby tensor for orthotropic matrix**

$$
\begin{aligned}
S_{IJ} &= \delta_{IJ}\frac{R^3}{2}\int_0^\infty \frac{(R^2+sk_I)^{-1}k_I}{\sqrt{\prod_{\ell=1}^3(R^2+sk_\ell)}}ds \\
&= \delta_{IJ}\frac{R^3}{2}\int_0^\infty \frac{k_I}{(R^2+sk_I)\sqrt{\prod_{\ell=1}^3(R^2+sk_\ell)}}ds \\
&= \delta_{IJ}\frac{R^3}{2}\int_0^\infty \frac{k_I}{R^2\left(1+\frac{sk_I}{R^2}\right)\sqrt{\prod_{\ell=1}^3 R^2\left(1+\frac{sk_\ell}{R^2}\right)}}ds \\
&= \delta_{IJ}\frac{R^3}{2}\int_0^\infty \frac{k_I}{R^2\left(1+\frac{sk_I}{R^2}\right)\sqrt{\prod_{\ell=1}^3 R^2\left(1+\frac{sk_\ell}{R^2}\right)}}ds \\
&= \delta_{IJ}\frac{1}{2}\int_0^\infty \frac{1}{\left(1+\frac{sk_I}{R^2}\right)\sqrt{\prod_{\ell=1}^3\left(1+\frac{sk_\ell}{R^2}\right)}}\cdot\frac{k_I}{R^2}ds \\
&= \delta_{IJ}\frac{1}{2}\int_0^\infty \frac{1}{\left(1+\frac{sk_I}{R^2}\right)\sqrt{\prod_{\ell=1}^3\frac{k_\ell}{k_I}\left(\frac{k_I}{k_\ell}+\frac{sk_I}{R^2}\right)}}\cdot\frac{k_I}{R^2}ds \\
&= \delta_{IJ}\frac{1}{2}\int_0^\infty \frac{1}{\left(1+\frac{sk_I}{R^2}\right)\sqrt{\prod_{\ell=1}^3\frac{k_\ell}{k_I}}\sqrt{\prod_{\ell=1}^3\left(\frac{k_I}{k_\ell}+\frac{sk_I}{R^2}\right)}}\cdot\frac{k_I}{R^2}ds \\
&= \delta_{IJ}\frac{1}{2}\sqrt{\prod_{\ell=1}^3\frac{k_I}{k_\ell}}\int_0^\infty \frac{1}{\left(1+\frac{sk_I}{R^2}\right)\sqrt{\prod_{\ell=1}^3\left(\frac{k_I}{k_\ell}+\frac{sk_I}{R^2}\right)}}\cdot\frac{k_I}{R^2}ds \\
&= \delta_{IJ}\frac{1}{2}\frac{\sqrt{\prod_{\ell=1}^3 k_I}}{\sqrt{\prod_{\ell=1}^3 k_\ell}}\int_0^\infty \frac{1}{\left(1+\frac{sk_I}{R^2}\right)\sqrt{\prod_{\ell=1}^3\left(\frac{k_I}{k_\ell}+\frac{sk_I}{R^2}\right)}}\cdot\frac{k_I}{R^2}ds \\
&= \delta_{IJ}\frac{1}{2}\frac{k_I^{3/2}}{\sqrt{\det K_0}}\int_0^\infty \frac{1}{\left(1+\frac{sk_I}{R^2}\right)\sqrt{\prod_{\ell=1}^3\left(\frac{k_I}{k_\ell}+\frac{sk_I}{R^2}\right)}}\cdot\frac{k_I}{R^2}ds \\
&= \delta_{IJ}\frac{1}{2}\frac{k_I^{3/2}}{\sqrt{\det K_0}}\int_0^\infty \frac{1}{(1+s')\sqrt{\prod_{\ell=1}^3\left(\frac{k_I}{k_\ell}+s'\right)}}ds'
\end{aligned}
$$

Here we substitute the variable $s$ as a new integral variable $s' = sk_I/R^2$ and the elliptic integral can be obtained from mathematica[2].

**Supplementary Note 3: Eshelby tensor for transversely isotropic matrix**

The Eshelby tensor for a particle inclusion in an orthotropic matrix with two anisotropy factors, $A = k_I/k_L$ and $B = k_I/k_M$. Suppose that $A \neq B$. Then, it can be derived as follows,

$$S_{IJ} = \delta_{IJ} \frac{(k_I)^{3/2}}{2\sqrt{\det(\boldsymbol{K}_0)}} \int_0^\infty \frac{1}{(s'+1)^{3/2}\sqrt{s'+A}\sqrt{s'+B}} ds'$$

$$= \delta_{IJ} \frac{(k_I)^{3/2}}{2\sqrt{\det(\boldsymbol{K}_0)}} \int_0^\infty \frac{1}{(s'+1)^{3/2}} \cdot \frac{1}{A-B} \left( \sqrt{\frac{s'+A}{s'+B}} - \sqrt{\frac{s'+B}{s'+A}} \right) ds'$$

$$= \delta_{IJ} \frac{(k_I)^{3/2}}{2\sqrt{\det(\boldsymbol{K}_0)}} \frac{1}{A-B} \left( \int_0^\infty \frac{\sqrt{s'+A}}{(s'+1)^{3/2}\sqrt{s'+B}} ds' - \int_0^\infty \frac{\sqrt{s'+B}}{(s'+1)^{3/2}\sqrt{s'+A}} ds' \right)$$

$$= \delta_{IJ} \frac{(k_I)^{3/2}}{2\sqrt{\det(\boldsymbol{K}_0)}} \frac{2}{A-B} \left( \frac{\mathrm{E}\left(\cos^{-1}(\sqrt{B}), \frac{A-1}{B-1}\right)}{\sqrt{1-B}} - \frac{\mathrm{E}\left(\cos^{-1}(\sqrt{A}), \frac{B-1}{A-1}\right)}{\sqrt{1-A}} \right)$$

where $E(\theta, m) = \int_0^\theta \sqrt{1 - m\sin^2\theta'}\, d\theta'$ is an elliptic integral of 2nd kind.

When $A = B$, the only remaining anisotropy factor of the matrix is $A = k_1/k_2$. Then, we have

$$S_{11} = \frac{A}{2} \int_0^\infty \frac{1}{(s'+1)^{\frac{3}{2}}(s'+A)} ds' = A \cdot \left( \frac{1}{A-1} - \frac{\sec^{-1}\sqrt{A}}{(A-1)^{3/2}} \right)$$

and

$$S_{22} = \frac{1}{2\sqrt{A}} \int_0^\infty \frac{1}{(s'+1)^2\sqrt{s'+\frac{1}{A}}} ds' = \frac{A}{2} \cdot \left( \frac{\sec^{-1}\sqrt{A}}{(A-1)^{1.5}} - \frac{1}{(A \cdot (A-1))} \right)$$

We note that under the results of the assumption of $A \neq B$, by taking the limit such that $B \to A$, we get the same result obtained from the assumption of $A = B$. We also note that, for the $A = B$ case, in the limit of $k_1 \to k_2 (A \to 1)$, the Eshelby tensor reduces to the isotropic matrix result, $S_{ij} = \frac{1}{3}\delta_{ij}$. The two independent values ($S_{11}$, $S_{22} = S_{33}$) are plotted in terms of $k_1/k_2$ in the Fig. S1(d) where we validate our solutions against the numerical evaluation of Eq. (4).

**Supplementary Note 4: Non-uniform heat flux within ellipsoidal inclusion having Kapitza's thermal resistance**

We simulate a single inhomogeneity problem with an ellipsoidal inclusion which has an aspect ratio of 2. The length of the cube edge is 10 m and the axis lengths of the inclusion are 1 m and 2 m respectively. The material properties used in this simulation are 1 W/mK (matrix) and 10 W/mK (inclusion). For the case of zero interfacial thermal resistance, the heat flux within the ellipsoidal inclusion is uniform (see Fig. S2). However, the heat flux is not uniform in the presence of an interfacial thermal resistance (see Fig. S2). The heat flux within the inclusion has a maximum at the centre of the inclusion and it decreases as the interfacial thermal resistance increases.

To explain the non-uniformity of the heat flux, we numerically calculate $M_{1111}$ for the ellipsoidal inclusion, assuming uniform heat flux within the inclusion. $M_{1111}$ is uniform when the shape of inclusion is a sphere, so the assumption is reasonable. However, when the inclusion shape is ellipsoidal, the value depends on the position in the inclusion, which means that the assumption cannot be used for the ellipsoidal inclusion case (see Fig. S3). Since the heat flux within the ellipsoidal inclusion is not uniform when interfacial thermal resistance exists, the conventional method cannot be used to calculate the localization tensor and effective modulus analytically

**Supplementary Note 5: Proof for Eq. (17)**

After using the divergence theorem at Eq. (15), we have

$$M_{ijms}(\mathbf{x}) = \frac{1}{R}\left[\frac{\partial N_{ijs}(\mathbf{x})}{\partial x_m} + \delta_{is} D_{mj}(\mathbf{x})\right],$$

where

$$N_{ijs}(\mathbf{x}) := \int_V \frac{\partial^2 G(\mathbf{y}-\mathbf{x})}{\partial y_i \partial y_j} y_s d\mathbf{y}$$

and

$$D_{mj}(\boldsymbol{x}) := \frac{\partial}{\partial x_m}\int_V \frac{\partial}{\partial y_j} G(\mathbf{y}-\mathbf{x}) d\mathbf{y}$$

with

$$G(\mathbf{x}-\mathbf{y}) := \frac{1}{4\pi\sqrt{\det(\boldsymbol{K}_0)[(\mathbf{x}-\mathbf{y})^T \boldsymbol{K}_0^{-1}(\mathbf{x}-\mathbf{y})]}}.$$

Since

$$\frac{\partial G(\mathbf{x}-\mathbf{y})}{\partial x_i} = -\frac{\partial G(\mathbf{x}-\mathbf{y})}{\partial y_i},$$

we get

$$N_{ijs}(\mathbf{x}) = \int_V \frac{\partial^2 G(\mathbf{y}-\mathbf{x})}{\partial y_i \partial y_j} y_s d\mathbf{y} = \frac{\partial^2}{\partial x_i \partial x_j}\int_V G(\mathbf{y}-\mathbf{x}) y_s d\mathbf{y} \ .$$

Our goal is to show

$$(K_0)_{ij}\frac{\partial N_{ijs}(\mathbf{x})}{\partial x_m} = \delta_{ms}.$$

Noting

$$K_0 = \mathrm{diag}(k_1, k_2, k_3),$$

we rewrite $G(\mathbf{y}-\mathbf{x})$ such that

$$G(\mathbf{y}-\mathbf{x}) = \frac{1}{4\pi\sqrt{k_1 k_2 k_3}} \cdot \frac{1}{\sqrt{\sum_{\ell=1}^{3}\frac{(x_\ell - y_\ell)^2}{k_\ell}}}.$$

This implies that

$$\frac{\partial N_{ijs}(\mathbf{x})}{\partial x_m} = \frac{\partial^3}{\partial x_m \partial x_i \partial x_j}\int_V G(\mathbf{y}-\mathbf{x}) y_s d\mathbf{y}$$

$$= \frac{1}{4\pi\sqrt{k_1 k_2 k_3}}\frac{\partial^3}{\partial x_m \partial x_i \partial x_j}\int_V \frac{y_s}{\sqrt{\sum_{\ell=1}^{3}\frac{(x_\ell - y_\ell)^2}{k_\ell}}} d\mathbf{y}.$$

Now we can get the following by substituting $\tilde{x}_\ell = \frac{x_\ell}{\sqrt{k_\ell}}$ and $\tilde{y}_\ell = \frac{y_\ell}{\sqrt{k_\ell}}$ :

$$
\begin{aligned}
&\frac{1}{4\pi\sqrt{k_1k_2k_3}}\frac{\partial^3}{\partial x_m\partial x_i\partial x_j}\int_V \frac{y_s}{\sqrt{\sum_{\ell=1}^{3}\frac{(x_\ell-y_\ell)^2}{k_\ell}}}d\mathbf{y}\\
&=\frac{1}{4\pi\sqrt{k_1k_2k_3}}\sqrt{k_S}\frac{1}{\sqrt{k_Mk_Ik_J}}\ \frac{\partial^3}{\partial\tilde{x}_M\partial\tilde{x}_I\partial\tilde{x}_J}\int_V \frac{\tilde{y}_S}{\sqrt{\sum_{\ell=1}^{3}(\tilde{x}_\ell-\tilde{y}_\ell)^2}}\cdot\frac{\partial\mathbf{y}}{\partial\tilde{\mathbf{y}}}d\tilde{\mathbf{y}}\\
&=\frac{1}{4\pi\sqrt{k_1k_2k_3}}\frac{\sqrt{k_S}}{\sqrt{k_Mk_Ik_J}}\frac{\partial^3}{\partial\tilde{x}_M\partial\tilde{x}_I\partial\tilde{x}_J}\int_{\tilde{\omega}} \frac{\tilde{y}_S}{\sqrt{\sum_{\ell=1}^{3}(\tilde{x}_\ell-\tilde{y}_\ell)^2}}\sqrt{k_1k_2k_3}d\tilde{\mathbf{y}}\\
&=\frac{\sqrt{k_S}}{4\pi\sqrt{k_Mk_Ik_J}}\frac{\partial^3}{\partial\tilde{x}_M\partial\tilde{x}_I\partial\tilde{x}_J}\int_{\tilde{\omega}} \frac{\tilde{y}_S}{\sqrt{\sum_{\ell=1}^{3}(\tilde{x}_\ell-\tilde{y}_\ell)^2}}d\tilde{\mathbf{y}}.
\end{aligned}
$$

This gives us that

$$
\begin{aligned}
&\sum_{I,J}(K_0)_{IJ}\frac{\sqrt{k_S}}{4\pi\sqrt{k_Mk_Ik_J}}\frac{\partial^3}{\partial\tilde{x}_M\partial\tilde{x}_I\partial\tilde{x}_J}\int_{\tilde{\omega}} \frac{\tilde{y}_S}{\sqrt{\sum_{\ell=1}^{3}(\tilde{x}_\ell-\tilde{y}_\ell)^2}}d\tilde{\mathbf{y}}\\
&=\sum_{I,J}k_I\delta_{IJ}\frac{\sqrt{k_S}}{4\pi\sqrt{k_Mk_Ik_J}}\frac{\partial^3}{\partial\tilde{x}_M\partial\tilde{x}_I\partial\tilde{x}_J}\int_{\tilde{\omega}} \frac{\tilde{y}_S}{\sqrt{\sum_{\ell=1}^{3}(\tilde{x}_\ell-\tilde{y}_\ell)^2}}d\tilde{\mathbf{y}}\\
&=\sum_{I}\frac{\sqrt{k_S}}{4\pi\sqrt{k_M}}\frac{\partial^3}{\partial\tilde{x}_M(\partial\tilde{x}_I)^2}\int_{\tilde{\omega}} \frac{\tilde{y}_S}{\sqrt{\sum_{\ell=1}^{3}(\tilde{x}_\ell-\tilde{y}_\ell)^2}}d\tilde{\mathbf{y}}\\
&=\sqrt{\frac{k_S}{k_M}}\frac{\partial}{\partial\tilde{x}_M}\int_{\tilde{\omega}} \Delta_{\tilde{x}}\frac{1}{4\pi\sqrt{\sum_{\ell=1}^{3}(\tilde{x}_\ell-\tilde{y}_\ell)^2}}\tilde{y}_Sd\tilde{\mathbf{y}}\\
&=-\sqrt{\frac{k_S}{k_M}}\frac{\partial}{\partial\tilde{x}_M}\int_{\tilde{\omega}} \delta(\tilde{x}-\tilde{y})\tilde{y}_Sd\tilde{\mathbf{y}}\\
&=-\sqrt{\frac{k_S}{k_M}}\frac{\partial\tilde{x}_S}{\partial\tilde{x}_M}\\
&=-\delta_{ms}.
\end{aligned}
$$

Hence, our goal is proven.

Here the capital index has the same value with small index and the repeated capital index are not summed over.

From the definition of Eshelby tensor (Eq. 4),

$$D_{mj}(\boldsymbol{x})K_{0_{ij}} = \frac{\partial}{\partial x_j}\int_V \frac{\partial G(\boldsymbol{x}-\boldsymbol{y})}{\partial y_m} d\boldsymbol{y}\, K_{0_{ij}} = S_{im}$$

So,

$$K_{0_{ij}} M_{ijms}(\mathbf{x}) = \frac{1}{R} K_{0_{ij}} \left[\frac{\partial N_{ijs}(\mathbf{x})}{\partial x_m} + \delta_{is} D_{mj}(\mathbf{x})\right] = \frac{1}{R}(-\delta_{ms} + S_{ms})$$

which is equivalent to

$$\boldsymbol{K}_0 : \boldsymbol{M} = \frac{1}{R}(-\boldsymbol{I} + \boldsymbol{S}).$$